# CRISM south polar mapping: First Mars year of observations

Adrian J. Brown[*1], Wendy M. Calvin[2], Patrick C. McGuire[3], Scott L. Murchie[4]

[1] SETI Institute, 515 N. Whisman Rd, Mountain View, CA 94043, USA
[2] Geological Sciences, University of Nevada, Reno, NV, 89557, USA
[3] McDonnell Center for the Space Sciences,
Washington University in St. Louis, MO 63130, USA
[4] Johns Hopkins University Applied Physics Laboratory, Laurel, MD, 20723, USA

[*] corresponding author, email: abrown@seti.org

Corresponding author:
Adrian Brown
SETI Institute
515 N. Whisman Rd Mountain View, CA 94043
ph. 650 810 0223
fax. 650 968 5830
email. abrown@seti.org

Short running title: "CRISM South polar mapping"

# ABSTRACT

We report on mapping of the south polar region of Mars using data from the Compact Reconnaissance Imaging Spectrometer for Mars (CRISM) instrument. Our observations have led to the following discoveries:

1. Water ice is present in the form of pole-circling clouds originating from the circum-Hellas region, beginning prior to $L_s$=162 and diminishing markedly at $L_s$=200-204.

2. It has previously been inferred by temperature measurements (Titus et al., 2003) and $CO_2$-$H_2O$ mixture spectral models (Langevin et al., 2007) that surface water ice was present in the Cryptic Region in the final stages of sublimation. The high resolution of CRISM has revealed regions where only water ice is present (not a $CO_2$-$H_2O$ ice mixture). This water ice disappears completely by $L_s$=252 and may be the source of water vapor observed by CRISM in southern latitudes between $L_s$=240-260 (Smith, et al., this issue).

3. We have estimated surface $CO_2$ ice grain size distributions for the South Pole Residual Cap (SPRC) and the seasonal $CO_2$ ice cap that covers it throughout summer spring and summer. Our analysis suggests that grain sizes peak at $L_s$=191-199 with an apparent grain size of ~7±1cm. By the end of the summer period our analysis demonstrates minimum apparent grain sizes of ~5±1mm predominate in the SPRC.

4. Fine grained $CO_2$ ice condenses from $L_s$=0-40, and extends symmetrically away from the geographic pole, extending beyond 80°S by $L_s$=4-10. No evidence for unusual $CO_2$ depositional processes in the Cryptic Region is observed up to $L_s$=16.

## KEYWORDS

# INTRODUCTION

U nderstanding the seasonal distribution of icy volatiles in the Martian polar regions is key to understanding the thermophysical balance of the Red Planet. The Viking Landers measured pressure variations of over 25% over a Martian year due to the sublimation and condensation of the polar caps (Hess et al., 1977; 1979; 1980). The pressure variations measured by the Viking Landers have been used to test heat budget models for the polar. regions (Wood and Paige, 1992; Guo et al., 2009). Increasingly detailed sublimation timing information from the Thermal Emission Spectrometer (TES) instrument (Kieffer et al., 2000) and composition information from the THEMIS and OMEGA spectrometers (Titus, 2005b; Langevin et al., 2007; Piqueux et al., 2008) has led to greater appreciation for the complexity of the polar sublimation/condensation cycle.

This study extends the compositional investigation of the south polar sublimation cycle by presenting seasonal maps of icy volatiles, as observed by the CRISM on the Mars Reconnaissance Orbiter (MRO) spacecraft during its first Mars year in orbit. We also present contemporaneous mosaics of Mars Color Imager (MARCI) camera data.

In this study, we refer to timings using the terms MY for Mars Year, specified as starting on 11 April 1955 by Clancy et al. (2000). MY 28 began 22 Jan 2006, MY 29 began 10 December 2007. The term "$L_s$", or solar longitude, refers to the celestial location of the sun when viewed from Mars (0-360°). This is a convenient measure of the Martian seasons – 180-360° is southern spring and summer. Aphelion occurs at $L_s$=71° and perihelion occurs at $L_s$=251°.

Within this paper, we use the terms 'grain size' and 'equivalent path length' interchangeably – CRISM data cannot be used to derive actual grain sizes, but comparison of CRISM spectra with spectra generated by models can be used to deduce the average path length of light within a material before a scattering event, and this is used to infer a grain size by assuming scattering takes place at grain boundary intervals (Bohren, 1987).

## CRISM

CRISM is a visible to near-infrared spectrometer sensitive to light with wavelengths from ~0.4 to ~4.0µm (Murchie et al., 2007). CRISM is ideal for detecting hydrated minerals and ices (Brown et al., 2008a) because it covers unique spectral absorption bands in the 1-4 micron region. Previous thermal infrared observations of the Martian polar regions have been able to infer ice composition by the temperature of the surface (e.g. TES and THEMIS), however CRISM is able to determine ice composition by the presence of absorption bands. This enables the detection of pure phases and mixtures of $CO_2$ and $H_2O$ ice. In addition, CRISM spectra can be used to derive ice grain sizes due to the concomitant changes in absorption band strength with grain size, as discussed further below.

CRISM is the highest spatial resolution spectrometer to observe the Martian polar regions to date. In high-resolution mode, CRISM's instantaneous field of view or pixel size corresponds to ~18.75m on the ground. CRISM has 640 pixels in the cross-track direction, which corresponds to 12km on the surface of Mars, however only 605 pixels see the surface (edge pixels are intentionally blocked as calibration fiducials) therefore the swath width exposes ~10.8km on the ground.

In mapping mode (used in this study) 10x spatial binning is employed on the spacecraft, before data comes back to Earth. A mapping swath therefore is 64 pixels wide, each pixel being ~187.5m in the cross-track direction. Along-track binning is controlled by exposure time, which is variable by the type of observation, to keep pixels approximately square.

**MARCI**

The MARs Color Imager (MARCI) camera is a super wide angle, fish eye lens instrument with a 1024 pixels-wide CCD. Each Martian sol, MARCI obtains 12 continuous terminator-to-terminator, limb-to-limb color images, covering 60° of longitude. MARCI's seven bandpass filters are positioned at 260, 320 (UV), 425, 550, 600, 650 and 725nm(VIS). The UV channels have 7-8km resolution and the VIS channels just under 1km resolution (Malin et al., 2001).

# METHODS

**CRISM Seasonal mosaics**

The seasonal mosaics discussed in this paper were constructed by collecting all the CRISM MSP (MultiSpectral Polar, which is a multispectral mapping mode for the polar regions) images acquired that extended south of 55° latitude for each two week MRO planning sequence.

In mapping mode, to keep data rates manageable, only a subset of the CRISM spectral bands are returned to Earth. MSP images have 19 bands in the CRISM 'S' (short wavelength, 0.36-1.0$\mu$m) and 55 bands in the CRISM 'L' (long wavelength, 1.0-3.9$\mu$m) detectors. In this paper, we have only used data from the 'L' detector, since it covers the absorption bands we require for mapping icy volatiles.

The images were mosaiced together using the "MR PRISM" software suite (Brown and Storrie-Lombardi, 2006) into a south polar stereographic projection with central meridian (longitudinal axis pointing directly down) of 180°, at a resolution of 1000x1000 pixels (Figure 1). Each successive image that overlapped a previous image overwrote the previous data. No averaging of data took place during mosaic contruction – the last pixel that was acquired in a region was the one that was used. Thus all the spectra in each prepared mosaic are unmodified spectra that were acquired at ~187.5m per pixel resolution.

MSP observations were acquired on a best efforts basis, and conflicting requirements such as high data loads from other parts of the planet, solar conjunction and CRISM cooler swaps dictated that a variable number of MSP observations of the polar region were acquired in each two week period. The summary of south polar mapping observations for each MRO planning period is presented in Table 1.

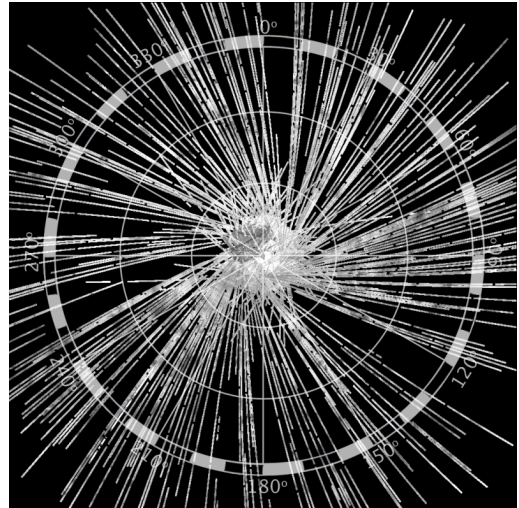

Figure 1. South polar mosaic of CRISM MSP images from Ls=295-003.

| MRO Planning Cycle | DOY (2007/08) | (MY) and Ls Range | S channel observations | L channel observations |
|---|---|---|---|---|
| 6 | (07)6-16 | (28) 161.9-167.3 | 56 | 66 |
| 7 | 17-30 | 167.9-175.0 | 147 | 168 |
| 8 | 31-44 | 175.5-182.8 | 231 | 253 |
| 9 | 45-58 | 183.4-190.9 | 370 | 376 |
| 10 | 59-72 | 191.5-199.2 | 455 | 463 |
| 11 | 81-86 | 204.6-207.6 | 201 | 201 |
| 12 | 89-100 | 209.4-216.1 | 446 | 447 |
| 13 | 101-114 | 216.7-224.8 | 475 | 434 |
| 14 | 115-116 | 225.4-226.0 | 87 | 91 |
| 15 | 132-142 | 236.1-242.4 | 404 | 428 |
| 16 | 143-156 | 243.1-251.3 | 464 | 428 |
| 17 | 157-168 | 251.9-258.9 | 256 | 258 |
| 18 | 172-181 | 261.5-267.1 | 234 | 232 |
| 19 | 188-198 | 271.5-277.8 | 346 | 346 |
| 20 | 199-212 | 278.4-286.5 | 199 | 200 |
| 21 | 213-219 | 287.1-290.8 | 123 | 125 |
| 22 | 227-240 | 295.6-303.4 | 119 | 118 |
| 23 | 241-254 | 304.0-311.6 | 146 | 146 |
| 24 | 255-268 | 312.2-319.7 | 210 | 210 |
| 25 | 269-282 | 320.2-327.5 | 180 | 180 |
| 26 | 283-296 | 328.1-335.2 | 172 | 173 |
| 27 | 297-309 | 335.7-342.1 | 152 | 150 |
| 28* | 350-352 | (29) 3.1-4.1 | 26 | 26 |
| 29* | 353-001 | 4.6-10.9 | 183 | 183 |
| 30* | (08)002-012 | 11.4-16.2 | 60 | 60 |
| 31* | 016-026 | 18.1-22.8 | 35 | 35 |
| 32* | 034-042 | 26.5-30.2 | 19 | 23 |
| 33* | 044-044 | 31.1 | 2 | 2 |
| 34* | 069-071 | 41.9-42.8 | 10 | 8 |

Table 1 - Number of CRISM MSP Strips taken of Mars southern pole (defined as all strips partially or completely poleward of 55°S) per fortnight (starting at cycle 6 of the MRO primary science mission) to July 17, 2007. Short channel is VNIR (0.4-1.0 μm). L channel is IR (1-4 μm). $L_s$ = Deg of solar longitude. Southern spring starts at $L_s$ = 180 and ends at $L_s$ = 270. After (Brown et al., 2007). MY = Mars Year, $L_s$ = solar longitude from Mars. MRO Planning cycles marked with an asterisk (*) are nominal - due to patchy coverage, we have amalgamated some cycles and omitted some.

All multispectral mapping images are acquired while CRISM is fixed in the nadir pointing direction.

The combination of field of view and instrumental constraints leads to the 'spaghetti strand' appearance of the resulting CRISM mosaics that are presented in this paper, with numerous gaps in surface coverage.

CRISM commenced imaging of the south polar region in January 2007, two months into the MRO mapping mission. This late start was due to the fact that CRISM requires sunlight to image the surface and until that time, the sun had not risen over the south polar region. CRISM multispectral mapping of the south pole was severely curtailed in mid-southern summer ($L_s$=309, MY 28) and finally halted on MY29 $L_s$=16 due to insufficient polar illumination.

**Reduction of radiance to apparent reflectance**

CRISM measures spectral radiance at sensor in $W.m^{-2}.sr^{-1}.\mu m^{-1}$. In order to retrieve the radiance value for each pixel, instrumental artifacts such as detector bias, electronic artifacts (bad pixels, ghosts, detector non-linearity), background subtraction (dark current), and stray light corrections are removed in the CRISM data processing pipeline (Murchie et al., 2007).

CRISM data is delivered by the instrument team to the Planetary Data System as radiance or I/F (radiance detected by the sensor ratioed to collimated flux power incident on the surface). Radiance at sensor is converted to I/F values by dividing the detected radiance by the solar flux at Mars (adopting symbology from Schott (1996)):

$$I/F = \frac{L_\lambda}{\left(\frac{\phi_\lambda}{\pi.(r_{MARS})^2}\right)}$$ (Eq.1)

where $L_\lambda$ is the spectral radiance at sensor in $Wm^{-2}sr^{-1}\mu m^{-1}$, $\Phi_\lambda$ is the spectral solar flux at 1 astronomical unit (AU) in $Wm^{-2}\mu m^{-1}$ and $r_{MARS}$ is the Mars-Sun distance in AU.

In order to retrieve apparent reflectance from the I/F data, we use the Digital Data Record supplied by the CRISM team to extract the solar incidence angle for each pixel in an observation and divide by the cosine of the incidence angle:

$$R = \frac{I/F}{\mu_0}$$ (Eq.2)

where $\mu_0$ =cos($i$), $i$ is the solar incidence angle, 0° when the sun is overhead and 90° at the horizon.

**Atmospheric effects**

We used the term 'apparent reflectance' (used here with the symbol *R*) to indicate that time-variable atmospheric effects (scattering from dust and ice aerosols) have not been accounted for. This quantity is also sometimes termed 'lambert albedo' since it assumes isotropic scattering from the surface, however we prefer 'apparent reflectance' to emphasize no atmospheric correction has been applied to this data. All data used in this paper utilizes apparent reflectance values.

In the polar regions, over bright icy regions (surface albedos > 0.5), the effect of the atmosphere is minimal, due to the extremely thin nature of the Martian atmosphere and the strong reflectivity of $CO_2$ ice, particularly in the 1-4 $\mu$m region, and of $H_2O$ ice, from 1-1.5$\mu$m. Most of the ice identifications discussed in this paper are over bright, large grain $CO_2$ ice regions therefore the absence of an atmospheric correction does not prevent the identification of ices and the determination of $CO_2$ ice grain size.

There are two important cases where the atmosphere has an effect on the analysis discussed below. Those are the determination of the presence of clouds, and the variable opacity of dust aerosols. These are discussed as they are encountered in the relevant parts of the text.

**$CO_2$ Ice Detection and Grain size Estimation Strategy**

In order to construct the estimated $CO_2$ grain size maps for the seasonal mosaics, we removed a linear continuum between two specified shoulder points and then fitted a single Gaussian shape using an iterative least squares method similar to that described in Brown (2006). We used the depth of the inverted Gaussian shape to evaluate the grain size by comparison with a spectral model.

We used two diagnostic absorption bands to determine the presence and abundance of $CO_2$ ice. These bands are centered at 1.435$\mu$m (shoulder points at 1.38 and 1.47$\mu$m) and 2.28$\mu$m (shoulder points at 2.2 and 2.4$\mu$m). The 1.435$\mu$m band was used to prepare maps of $CO_2$ ice presence, and the 2.28$\mu$m band was used to estimate the grain size. The 2.28$\mu$m band is able to give a more accurate grain size estimate because 1.) it is not obstructed by $CO_2$ atmospheric absorption bands and 2.) it is not overlapped by similar size $H_2O$ ice absorption bands (Calvin and Martin, 1994).

To link the depth of the absorption band to a grain size, we constructed a range of spectral models using the Shkuratov (1999) surface particulate spectral reflectance model. The Shkuratov surface model is a simple geometrical optics (ray tracing) model that converts optical constants into the expected reflectance of a particulate surface, given a monodisperse grain size distribution (all grains the same size) and porosity. The model is one-dimensional and treats the powder as a stack of slabs of width equal to the grain size, and with holes in the slabs to represent the porosity. It has been used previously to retrieve the grain size of ices in the Martian south polar region (Langevin et al., 2007). We used the optical constants of solid $CO_2$ ice as derived by Hansen (1997; 2005). All band depth calculations were conducted on spectra that had been convolved to the CRISM multispectral wavelengths. All mixtures were performed at the full resolution of the optical constants (roughly 0.2nm in the 1-4µm range) before convolution to CRISM multispectral wavelengths.

*$CO_2$ snowpack density and porosity estimation.* Several studies have derived the density of the Martian $CO_2$ seasonal deposits, with particular emphasis on the north pole seasonal deposits. Solid $CO_2$ has a density of 1606 kg m$^{-3}$. Smith et al. (2001) used combined MOLA gravity/elevation measurements to derive a density of 910+/-230kg.m$^{-3}$ for the north pole. Feldman et al. used the Gamma Ray Spectrometer (GRS) on Mars Odyssey to derive a density of 910 kg m$^{-3}$ (Feldman et al., 2003). Titus et al. suggest that the south polar seasonal cap density is probably around 1000 kg m$^{-3}$. Guided by these estimates, we adopt a porosity of 60% for our $CO_2$ ice spectral models. A constant density is unlikely to be realized in nature, but one value needs to be chosen for consistent results.

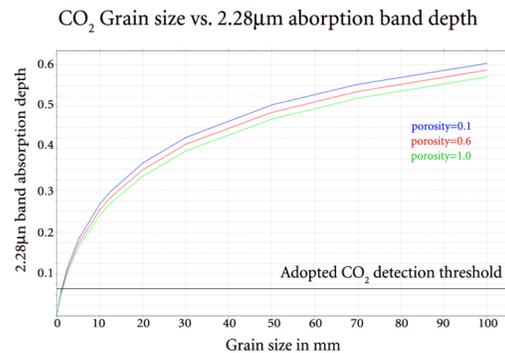

Figure 2.1b. Modeled relationship of particulate $CO_2$ ice 2.28 µm absorption band depth versus grain size for porosity of 0.1, 0.6 and 1.0. The adopted $CO_2$ detection threshold of 0.07 is shown.

In Figure 2.1a, we plot the depth of the 1.435µm absorption band against the modeled $CO_2$ ice grain size, using Shkuratov's surface reflectance method with an assumed porosity ('q' in Shkuratov's terms) of 0.1, 0.6 and 1.0.

In Figure 2.1b, we plot the relative band depth of the 2.28µm absorption band, also determined using Shkuratov's surface reflectance model for pure $CO_2$ ice. This figure presents the relationship we used to infer $CO_2$ grain sizes from CRISM reflectance measurements. To give an estimate of how the grain size retrieval is affected by Shkuratov's porosity parameter, we have also plotted the relative band depth vs. grain size for a porosity of 0.1 (extremely fluffy snow) and 1.0 (slab ice) bracketing the possible extremes we might encounter.

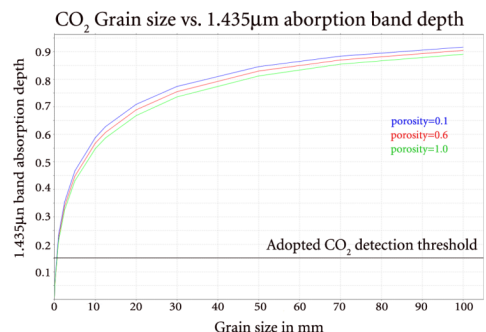

Figure 2.1a. Modeled relationship of particulate $CO_2$ ice 1.435 µm absorption band depth versus grain size for porosity = 0.1, 0.6 and 1.0. The adopted $CO_2$ detection threshold of 0.16 is shown.

For $CO_2$ ice with a grain size of 5mm, the variation between porosity extremes is ±1mm, for grain sizes of 30 mm, ±3mm and for grain size of 80mm, the variation is ±9mm. Therefore the unknown porosity of the ice introduces a relatively large (up to ±10%) uncertainty into our grain size derivation. This is the largest source of error in our calculations. CRISM SNR at 2.28µm is ~400 and contributes a worst case error for a 2.28µm band depth of 0.08 of (1/400)/0.07=3.5%. If we assume the porosity error and instrumental noise are independent errors, and convolve them according to *total error = sqrt(errorA$^2$+errorB$^2$)* then the worst case instrumental error adds ~0.6% to the porosity error. Therefore we use ±10% (rounded up to 1 significant digit) as our error estimate when quoting grain sizes in the Observations section below.

In order to create the interpretive ice maps for the seasonal mosaics, we used the 1.435µm $CO_2$ absorption band depth to 'detect' the presence of $CO_2$ ice. During testing with a range of threshold values, we found $CO_2$ gas absorption in the Martian south polar atmosphere tops out at relative band strength of ~0.15 at 1.435µm. Therefore, to detect $CO_2$ ice without spurious atmospheric detections, we chose to use a threshold value of 0.16 for the relative strength of this absorption, as indicated on Figure 2.1a. According to our models, this leads to a lower detection limit of pure $CO_2$ grains of approximately 750 microns in size. In comparison, the OMEGA team used a threshold of 0.2 (Langevin et al. (2007), para. [47]).

In order to estimate $CO_2$ grain sizes, the relative band depth at 2.28µm was used in conjunction with the optical model discussed previously (Figure 2.1b). We found that a lower threshold of 0.07 (shown as a horizontal line) eliminated non-$CO_2$ ice returns, which translates to a lower threshold of pure $CO_2$ grains about 1200 microns in size. This threshold was chosen to prevent noisy spectra being misidentified and put into the relative band depth histograms.

*Linear and intimate mixtures with soil and dust.* Grain sizes indicated by the mosaics will be inaccurate when $CO_2$ does not cover the entire pixel (~187.5m across) or if the $CO_2$ ice is not 'optically thick'. If $CO_2$ does not cover the entire region, the grain sizes will be underestimated. This effect will be most prevalent on the edges of the $CO_2$ cap. To illustrate the effect of impurities on the derivation of $CO_2$ ice grain size, we have prepared linear and intimate mixing models of $CO_2$ ice with palagonite, which is a 'dust' or 'dirt' representative that is often used for Martian soil (Brown et al., 2008a).

Linear mixing assumes that light reflect by one component does not interact with the other component before detection by the sensor. This approximates the situation where two components are side by side within one pixel. In the intimate mixing model, light reaching the sensor has bounced many times between the two components.

*Linear mixtures of CO₂ ice and soil*. We prepared a linear spectral model that point-by-point averages Shkuratov-derived reflectance spectra from two components – $CO_2$ ice with grain size of 2.5 mm and a porosity of 0.6 and palagonite with a grain size of 100 microns and porosity of 0.6. Figure 2.1c shows the results of employing our model with different linear mixture ratios, and calculating the 2.28 μm band depth to determine the grain size (by assuming the relationship in Figure 2.1b). As can be determined from the graph, the relationship between inferred grain size and mixing ratio is not linear, in fact, the partial coverage of the $CO_2$ ice must drop to less than 60% before the grain size estimate decreases to 80% of its true value.

*Intimately mixed dust and CO₂ ice*. The assumption of pure $CO_2$ ice is not likely to be realized on the south polar cap, since it is likely that small amounts of 'dust' or darker material will be intimately mixed in with the $CO_2$. For example, Doute et al. (2006) estimated that the 'typical' SPRC $CO_2$ ice had dust contents of 0.03-0.06% and were 99.9% $CO_2$ ice. It has been shown that even such a small amount of dust intimately mixed in with $CO_2$ ice will decrease the band depth and lead to a decrease in grain size estimation (Kieffer, 1990; Calvin and Martin, 1994). To illustrate the effect of intimate mixing on the derived grain sizes, we have used the Shkuratov coarse grain intimate mixing model (his equation 14) and the same $CO_2$ ice grains (2.5 mm, porosity=0.6) and palagonite grains (100 microns, porosity=0.6) as used in the linear mixing model, and plotted the results on Figure 2.1c. As can be seen, intimate mixtures have a much stronger effect on the albedo (and hence the inferred grain sizes) than linear mixing. Mixing $CO_2$ ice (2.5mm grain size) intimately with palagonite (100 microns grain size and a volume amount of 0.05) will result in the grain size of the $CO_2$ ice grain size being estimated at 1.8mm, a 30% error.

Our models of linear and intimate mixing of impurities with $CO_2$ ice show that impurities reduce the apparent grain size, and this is why we use the term 'minimum apparent' grain sizes when describing our findings.

*CO₂ Ice Clouds*. Assuming ice particles in the polar regions are not replenished and follow the Stokes-Cunningham settling model, 10μm grain sizes are likely to have an atmospheric residence time of around 10 hours, and therefore grains larger than 10μm are likely to fall to the surface in the thin Martian atmosphere (Conrath, 1975). Since persistent $CO_2$ ice clouds should have particle sizes of less that 10μm and our method of detecting $CO_2$ ice involves thresholds that eliminate sensitivity to very fine grained (<1mm) $CO_2$ grains, our $CO_2$ index maps are not sensitive to $CO_2$ ice clouds. $CO_2$ ice clouds have been suggested to form in the polar night when atmospheric temperatures are cold enough for $CO_2$ ice to condense (Hinson and Wilson, 2002). MOLA echoes during the polar night have been interpreted as returns from convective $CO_2$ ice clouds (Pettengill and Ford, 2000; Colaprete and Toon, 2002; Tobie et al., 2003). Recent SPICAM solar occultation observations have been interpreted to result from the presence of noctilucent $CO_2$ ice clouds at ~100km with grain sizes of ~100nm (Montmessin et al., 2006). If $CO_2$ ice clouds only form during the polar night, they will not be visible in the CRISM MSP dataset, which was collected when the surface is illuminated.

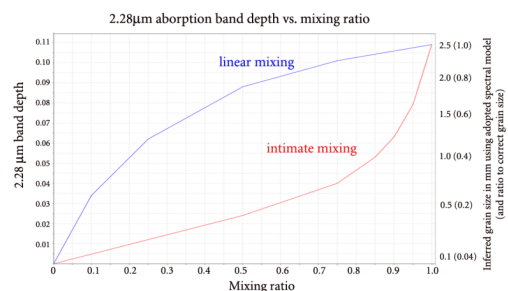

Figure 2.1c. Linear vs. intimate mixing. Modeled relationship of particulate $CO_2$ ice 2.28 μm absorption band depth versus mixing ratio when $CO_2$ ice of grain size 2.5mm is mixed with palagonite of grain size 100 microns. Porosity of all components is 0.6. Linear mixing is in blue, intimate mixing in red. Incorrectly inferred grain sizes using the standard spectral model used in this study are plotted on the right axis.

## $H_2O$ Ice Detection Strategy

In order to detect water ice, we used a water ice index similar to that suggested by Langevin et al. (2007), adjusted to accommodate CRISM multispectral mapping bands. The CRISM adjusted MSP $H_2O$ index formula is:

$$H_2O\,index = 1 - \frac{R(1.500)}{R(1.394)^{0.7}\,R(1.750)^{0.3}}$$  (Eq. 3)

where $R(\lambda)$ is the apparent reflectance at wavelength $\lambda$ in microns. Langevin et al. (2007) used wavelengths at 1.385 and 1.772µm in the denominator, and for comparison with their work we have approximated this relationship as closely as possible.

As for the $CO_2$ index threshold, we generated maps of water ice using a range of different thresholds and checked by hand how many incorrect or dubious water identifications were made for each threshold. We required that the $H_2O$ 1.5µm absorption band had to be readily apparent above CRISM noise. CRISM has a nominal SNR of ~400 in this region of the spectrum (Murchie et al., 2007) however the achieved SNR will be adversely affected by the low-light conditions in the south polar MSP dataset. We adopted a $H_2O$ index threshold of 0.125 since this value removed random noise from our $H_2O$ detection maps.

*Model to test $H_2O$ index threshold.* To estimate how much water ice is detectable with an adopted $H_2O$ index of 0.125, we created a spectral model of a $CO_2$ snowpack with embedded (intimately mixed) $H_2O$ grains. The grain sizes chosen have been guided by the results of previous research (Kieffer et al., 2000; Glenar et al., 2005; Doute et al., 2006; Langevin et al., 2007) but are only intended as a basic plausible model rather than a real physical situation. We used the Shkuratov (1999) intimate mixture particulate model (his equation 14) to generate approximate spectra of $CO_2$ mixed with water ice. We assumed a two component intimate mixture of porosity 0.6, with $CO_2$ ice grain size of 60000 microns and a water ice grain size of 500 microns (guided by the Doute et al. (2006) best fit to 'typical Bright residual cap') and varied the water amount by volume from 0 to 0.05. We then repeated this calculation series for $CO_2$ ice grain size of 2000 microns and a $H_2O$ ice grain size of 100 microns (more typical of what we observe in the seasonal cap or Cryptic Region). The mixing calculations were done at full spectral resolution of the optical constants and the $H_2O$ index calculated after convolution to MSP wavelengths. The results of this modeling are shown in Figure 2.2a.

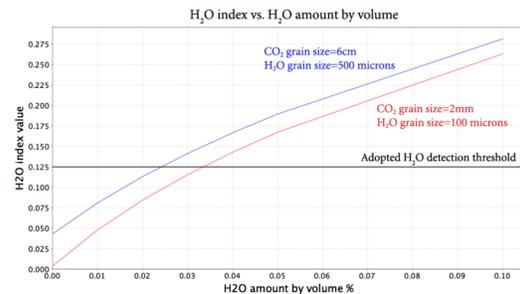

Figure 2.2a. Modeled relationship of particulate $H_2O$ ice index (Equation 3) versus modeled water amount by volume (in the presence of $CO_2$ grains 2mm in size).

This simple spectral model gives some idea of the sensitivity of the $H_2O$ index for plausible $H_2O$ and $CO_2$ mixtures. As can be seen in the figure, this simple model suggests that using our chosen threshold, for the chosen range of $CO_2$ and $H_2O$ ice grain sizes, the water ice will be detected when it is present in amounts greater than 2.5-3.5% by volume.

*$H_2O$ index Modeling assumptions*. The caveats discussed above for our $CO_2$ ice spectral modeling also apply to our $H_2O$ ice spectral modeling. Our modeling implicitly assumes $H_2O$ and $CO_2$ are intimately mixed on the surface, however as will be shown later, it is probable that some of the $H_2O$ ice we detect is in the form of water ice clouds, and our intimate mixture models are not expected to perform adequately when this situation exists. Another situation that our intimate mixing model will not simulate is cold trapped $H_2O$ molecules forming layers over a $CO_2$ substrate. In this layered configuration, it is likely that $CO_2$ absorption features will be suppressed even further than in the intimate mixing situation, and smaller amounts of $H_2O$ (i.e. even less than 2.5% as quoted above) may be detectable.

*$H_2O$ ice grain size estimation*. Langevin et al. (2007) suggested a ratio of the 3.4/3.525$\mu$m bands of OMEGA would enable the detection of water ice clouds, since they made the assumption that small water ice grain sizes correspond to water ice cloud (Langevin et al. (2007), para [32]). To use this method with our CRISM dataset, we created a similar index, using the 3.397$\mu$m and 3.503$\mu$m CRISM bands, since these are available in the MSP spectral subset.

$$H_2O\,grain\,Size\,Index = \frac{R(3.397)}{R(3.503)} \qquad\qquad\text{(Eq. 4)}$$

In order to justify the use of this ratio for detection of fine grained water ice, we created a simple, one component Shkuratov-based spectral model using $H_2O$ ice of porosity 0.6 and varying grain size. The results are shown in Figure 2.2b, for both the Langevin index and the slightly less effective modified CRISM index. It can be seen that the spectral factor behind the effectiveness of the 3.4/3.525μm ratio is the peak at 3.8μm. This peak is present in small grains (< 50 microns) but decreases in size and eventually disappears for grains larger than 50 microns. Langevin et al. (in para [32]) suggest the 3.4/3.525μm index might be used to discriminate between grains of mm to cm, however we contend its usefulness decreases markedly for grains larger than 50 microns. As Langevin et al. noted (in their para [32] and caption for their Figure 30), the index can be confused in the presence of large grained $CO_2$ ice due to overlapping of absorption bands. In addition, the SNR for CRISM in the 3-4μm region drops below 100 (Murchie, 2007). These two factors make interpretation of the $H_2O$ ice grain size ratio challenging.

Assuming, as discussed earlier for $CO_2$ ice clouds, that persistent water ice clouds are likely to have grain sizes around 10 microns and definitely smaller than 50 microns, therefore we consider a 3.397/3.503 ratio value of less than 0.6 to be due to fine grained ice, and potentially persistent water ice cloud.

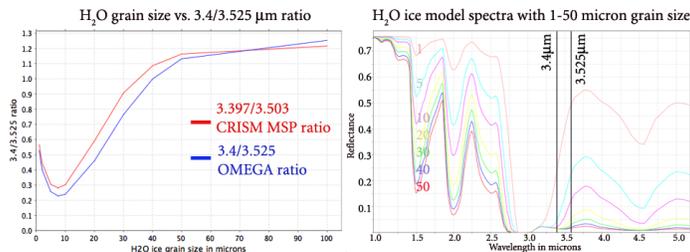

Figure 2.2b. (left) Plot of $H_2O$ ice grain size index (Equation 4) versus modeled water ice grain size and (right) model water ice spectra with grain sizes from 1-50 microns, vertical lines show the ratio points.

**CROCUS Line Mapping**

Despite the sometimes sparse CRISM mapping coverage of the polar regions it is possible to obtain an estimate of when the Cap Recession Observations indicated $CO_2$ has Ultimately Sublimated (CROCUS line - which corresponds to the retreating seasonal cap (Kieffer et al., 2000)) for comparison with previous studies (James et al., 1987; Kieffer et al., 2000; Titus, 2005a; Piqueux and Christensen, 2008).

In this study, we calculated the CROCUS line using the icy volatile mosaics. We defined the edge of the retreating cap as the location where $CO_2$ or $CO_2+H_2O$ was found, furthest away from the south pole. Due to the patchy nature of CRISM MSP coverage, we used a computer algorithm to automatically recognize the edge of the cap by laying down transects away from the pole, and finding the point along that transect where the last $CO_2$ or $CO_2+H_2O$ point was mapped (using the thresholds defined previously). In order to avoid outliers or spurious noise, for positive detection of the cap edge we required that the pixel in question be surrounded by five adjacent pixels that were also mapped as containing $CO_2$ or $CO_2+H_2O$ ice. We used linear interpolation to jump over areas of missing data. We used a variable lookahead scheme to look for the next point with the greatest radius away from the south pole.

**MARCI Daily mosaics**

The MRO orbit is 117 minutes long, and over the course of the 12.6 MRO orbits per Martian sol, MARCI maps out a set of swaths under the spacecraft ground track. These swaths are mosaiced at Malin Space Science Systems into a cylindrical projection "Daily Global Map" (DGM) product. We use this DGM product, re-projected to south polar stereographic orientation to 60°S latitude, and replacing the green channel with an average of red and blue to provide a better approximation of Martian surface color. The DGMs include a photometric correction averaging over the past several martian sols. This correction imparts a dark ring, coincident with the farthest outlier of bright ice material, particularly evident in the later $L_s$ images in Figure 3 (see e.g. $L_s$=260.2 in Figure 3.4 – the dark core is an artifact of the processing technique). These images are not accurate apparent radiance products, but provide confirmation at improved spatial fidelity of the seasonal ice mapping provided by CRISM. Camera acquisitions are automatically keyed to season so that data strips do not always cover the geometric pole and this results in dark gores at the center of the images at both early and late $L_s$ values. Obscuration by the MY28 large global dust event is particularly evident from $L_s$=270-304. Off-axis observations for other instruments also impart gores in the DGMs. These are observed as either missing data or areas along the mosaic seams where the color bands do not perfectly line up so that there are blue, orange or white color artifacts along some radial strip boundaries. For each CRISM time period we have selected the best corresponding DGM image with the fewest artifacts.

# OBSERVATIONS

**Seasonal mosaics**

Figure 3 shows the seasonal mosaics of CRISM, alongside MARCI data for taken within the same time period.

A variable number of CRISM MSP observations was taken per two week cycle, this number is indicated as '*n*' on the left hand side of the images. In the final Figure (Figure 3.7) since the observations are so few, these have been lumped into periods greater than the two week MRO planning cycle. Where the indicated period is less than two weeks, it has been shortened to include the period when the earliest and latest CRISM image was acquired in that cycle.

\<these figures are too large so they are available on the web at these links:\>
http://abrown.seti.org/research/downloads/brown-jgr-crism-sth-pole-my-28/figure3.1.png
Figure 3.1. South polar seasonal mosaics for period from MY 28 Ls=160-191. (left) Daily MARCI mosaic. (center) $CO_2$ 1.435 µm absorption depth (right) Ice identification maps

http://abrown.seti.org/research/downloads/brown-jgr-crism-sth-pole-my-28/figure3.2.png
Figure 3.2. South polar seasonal mosaics for period from MY 28 Ls=191-225. (left) Daily MARCI mosaic. (center) $CO_2$ 1.435 µm absorption depth (right) Ice identification maps

http://abrown.seti.org/research/downloads/brown-jgr-crism-sth-pole-my-28/figure3.3.png
Figure 3.3. South polar seasonal mosaics for period from MY 28 Ls=225-260. (left) Daily MARCI mosaic. (center) $CO_2$ 1.435 µm absorption depth (right) Ice identification maps

http://abrown.seti.org/research/downloads/brown-jgr-crism-sth-pole-my-28/figure3.4.png
Figure 3.4. South polar seasonal mosaics for period from MY 28 Ls=260-290. (left) Daily MARCI mosaic. (center) $CO_2$ 1.435 µm absorption depth (right) Ice identification maps

http://abrown.seti.org/research/downloads/brown-jgr-crism-sth-pole-my-28/figure3.5.png
Figure 3.5. South polar seasonal mosaics for period from MY 28 Ls=290-330. (left) Daily MARCI mosaic. (center) $CO_2$ 1.435 µm absorption depth (right) Ice identification maps

http://abrown.seti.org/research/downloads/brown-jgr-crism-sth-pole-my-28/figure3.6.png
Figure 3.6. South polar seasonal mosaics for period from MY 28 Ls=330 to MY29 Ls=10. (left) Daily MARCI mosaic. (center) $CO_2$ 1.435 µm absorption depth (right) Ice identification maps

http://abrown.seti.org/research/downloads/brown-jgr-crism-sth-pole-my-28/figure3.7.png
Figure 3.7. South polar seasonal mosaics for period from MY 28 $L_s$=10 to MY 29 $L_s$=40. (left) Daily MARCI mosaic. (center) $CO_2$ 1.435 µm absorption depth (right) Ice identification maps

Figure 4 shows representative spectra of different spectral classes at different points and times in the south polar cap taken from the MSP dataset. The spectral differences between $CO_2$ ice and gas absorptions can be appreciated by comparing Figure 4.a to Figure 4.f. The difference between fine grained and coarse grained $CO_2$ ice can be seen in Figure 4.a and 4.b, and for fine and coarse grain $H_2O$ ice comparison can be made between Figures 4.c and 4.d. Note that "fine" and "coarse grain" spectra for $CO_2$ ice and $H_2O$ ice correspond to physically different size grains.

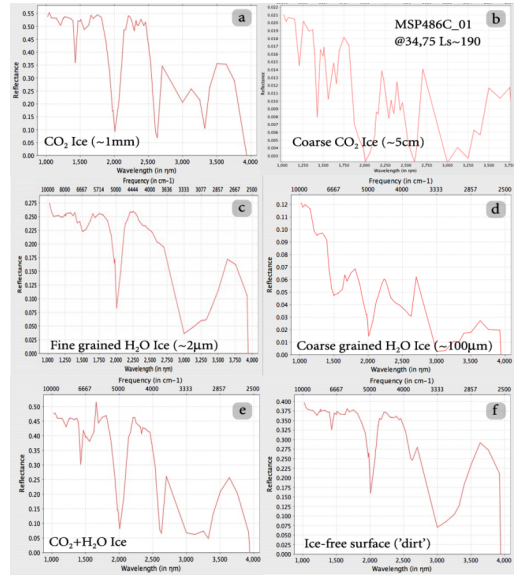

Figure 4. Representative spectra of spectral classes taken from the MSP dataset. See text for discussion. Reflectance here is "apparent reflectance" (incidence angle corrected) as discussed in the text and no atmospheric correction has been carried out. Note the subtle effect of $H_2O$ ice on $CO_2$ ice (in Figure 4.e there is a slight decrease in reflectance at 1.5µm relative to Figure 4.a)

**CROCUS lines and Cap Area**

Figure 5 plots the CROCUS line (defined here as the boundary of $CO_2$ ice as detected by the seasonal mosaics and determined by the threshold in figure 2.1) as a function of time. Due to our interpolation method which is required to skip areas of missing data, the edge of the cap is unrealistically simplified, however our results are similar to observations on previous years (James et al., 1979; James et al., 1987; Kieffer et al., 2000; Benson and James, 2005; Titus, 2005a; Giuranna et al., 2007; James et al., 2007; Piqueux and Christensen, 2008).

Using the polygons in Figure 5, we are able to estimate the surface area of the cap. We use a method of computing the area of an irregular polygon supplied by the publicly available Computational Geometry Algorithms Library (CGAL). Figure 6 plots the estimate of the area of the seasonal cap and the rate of change in the area of the cap per 24hrs versus solar longitude. The rate of change of the cap area averages around 52,000 km$^2$ per day from $L_s$=168 to $L_s$=310 and peaks at just under 110,000 km$^2$ per day at $L_s$=225.

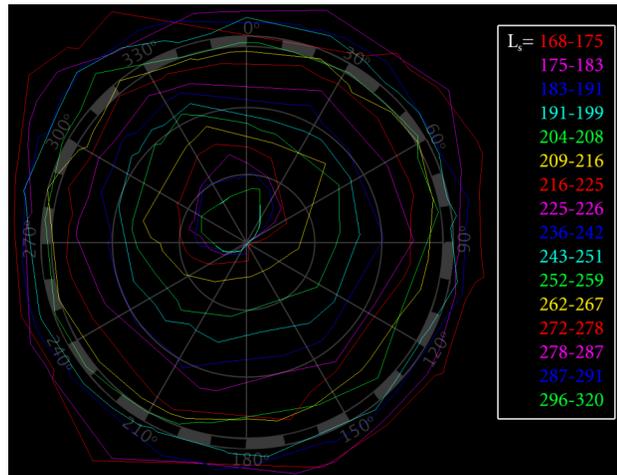

Figure 5. $CO_2$ seasonal cap edge (CROCUS) lines for MY28 $L_s$=168-320.

Assuming $CO_2$ mass load of 700 kg.m$^{-2}$, as measured by GRS (Feldman et al., 2003) during peak sublimation times at $L_s$=225, 7.7x10$^{13}$ kg of $CO_2$ is sublimed into the atmosphere per day.

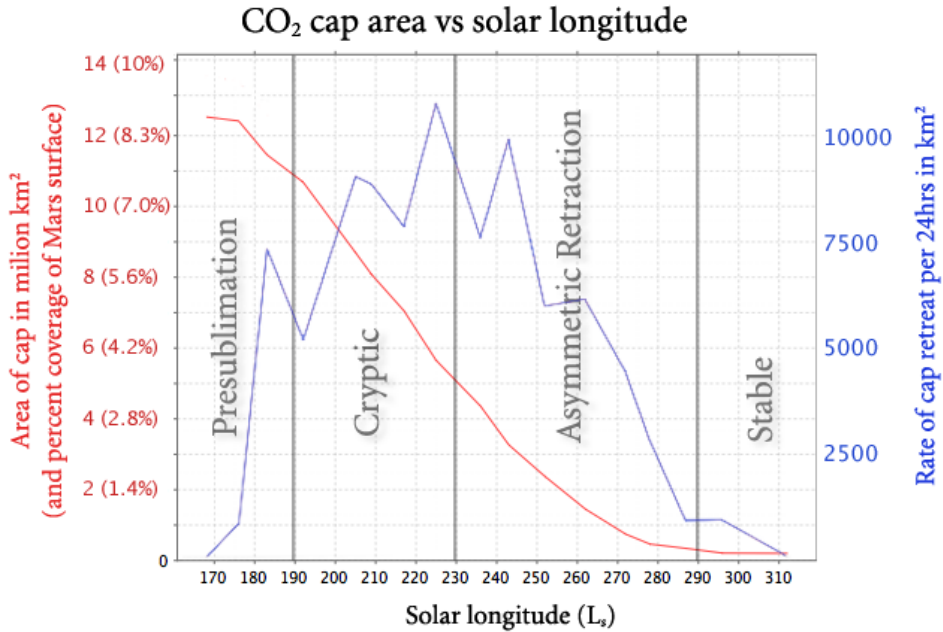

Figure 6. $CO_2$ seasonal cap area for $L_s$=168-312 (in red) and cap rate of retreat in km$^2$ per 24 hours (in blue). Sublimation phases indicated are discussed in the text.

**Histograms of seasonal $CO_2$ ice grain size**

Figures 7.1-7.5 show the seasonal histograms of derived $CO_2$ grain sizes derived from the CRISM mosaics.

The histograms have been constructed using the relative band strength of the 2.28μm $CO_2$ ice feature. All spectra that exhibited a relative band depth threshold of 0.07 were classed as $CO_2$ ice spectra, and included in the histogram data. All other spectra (not containing $CO_2$ ice) were omitted from the histograms. The spectra were binned into 1 of 100 bins according to their relative band depth. Each bin was summed to create the histogram, and then each bin was normalized according to the number of $CO_2$ ice pixels found in each two week mosaic.

The histograms include data from periods matching the MRO planning cycle, from $L_s$=162-016, corresponding to the mosaics in Figures 3.1-3.7. Each histogram has been grouped to reflect the different stages of the springtime recession (pre-sublimation, cryptic, asymmetric recession, stable and recondensation periods). Due to the small number of pixels available for analysis, we show only two histograms in the recondensation period ($L_s$=4.6-10.9 and 11.4-16.2).

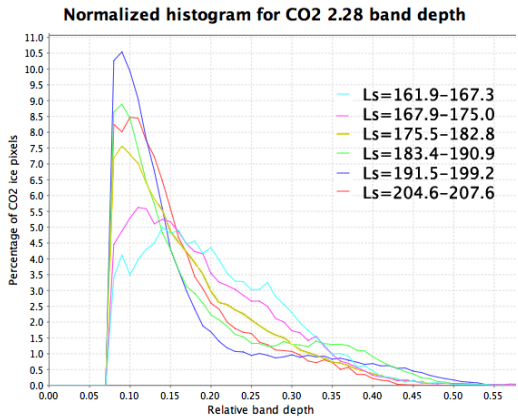

Figure 7.1 Histograms showing $CO_2$ surface ice grain size distributions for the Presublimation period corresponding to six MRO planning cycles used to prepare Figure 3 mosaics.

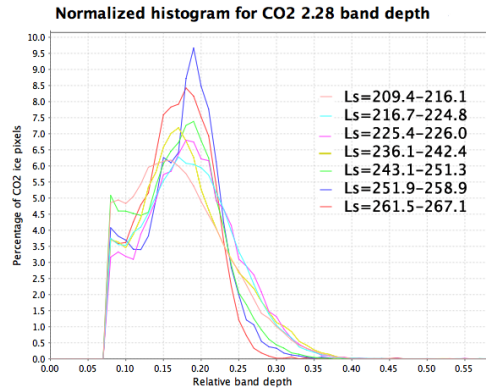

Figure 7.2 Histograms showing $CO_2$ surface ice grain size distributions for the Cryptic period corresponding to seven MRO planning cycles used to prepare Figure 3 mosaics.

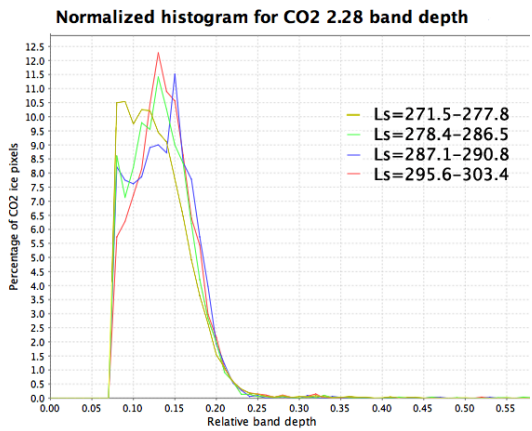

Figure 7.3 Histograms showing $CO_2$ surface ice grain size distributions for the asymmetric retraction period corresponding to four MRO planning cycles used to prepare Figure 3 mosaics.

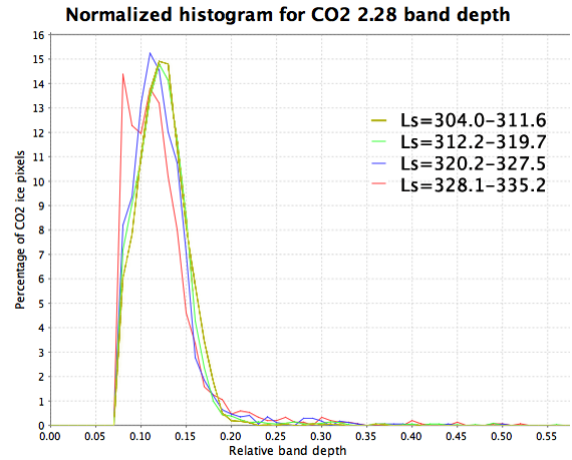

Figure 7.4 Histograms showing $CO_2$ surface ice grain size distributions for the stable period corresponding to four MRO planning cycles used to prepare Figure 3 mosaics

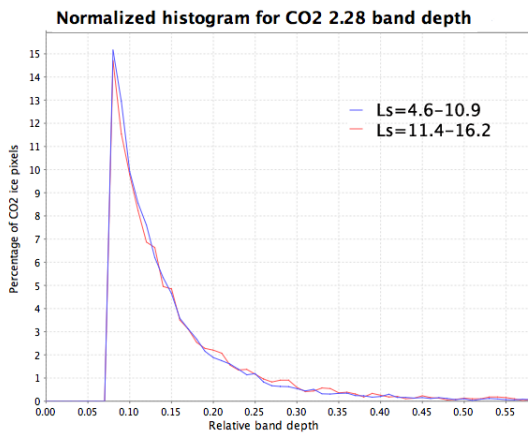

< Figure 7.5 Histograms showing $CO_2$ surface ice grain size distributions for the recondensation period corresponding to two MRO planning cycles used to prepare Figure 3 mosaics.

# DISCUSSION

**Interpretation of Seasonal Mosaics**

Based on the spatial extent and rate of sublimation or condensation, we divide our discussion of the seasonal mosaics into five major phases – presublimation, cryptic, retraction, stable and recondensation phases. The first phase, from $L_s$=160-200 we term *presublimation* phase. From $L_s$=200-240, is the appearance of the *cryptic* region, from $L_s$=240-290 is the cap *asymmetric retraction* phase and from $L_s$=290-340 is the *stable* cap phase. These phases are indicated on Figure 7. Following $L_s$=340, CRISM MSP coverage is spotty, though it is clear from Figure 7 that $CO_2$ has reappeared and extends further equatorward than 80°S at $L_s$=4-10. This $CO_2$ is most likely the growing surficial cap, as discussed in more depth below.

*Presublimation phase ($L_s$=160-200)*. Figures 3.1-3.2 cover this phase. The $CO_2$ seasonal cap clearly extends equatorward of 60°S, with relatively low, homogenous albedo and the cryptic region is not readily visible.

The edge of the $CO_2$ seasonal cap displays a decrease in the strength of the 1.435$\mu$m $CO_2$ absorption band as shown in the middle column of Figure 3.1 and 3.2. This decrease in $CO_2$ band strength could be due to a decrease in area covered by the $CO_2$ cap (at fractions of the ~187.5m scale) or due to a decrease in the grain size as the edge of the cap is sublimating. These two effects are not separable without high resolution images of the cap edge. The edge of the cap thus defined forms an annulus around 10-100km wide around the retreating cap.

*$H_2O$ ice clouds from $L_s$=160-200*. $H_2O$ ice is also present over the 60-80°S latitude range, encircling the geographic pole. These $H_2O$ ice pixels display low $H_2O$ ice grain index ratios (Eq. 4), therefore we interpret this water ice as polar vortex clouds originating from the Hellas region (Haberle et al., 1979). There is a large decrease in the area covered by water ice around $L_s$~200.

It is unclear what relationship exists between the water ice clouds that are dominant prior to $L_s$=200 and the subsequent weaker water ice detections on the seasonal cap, mostly in the cryptic region, after $L_s$=200. It is possible that the water ice clouds may condense on cryptic region after $L_s$=200, however it is also possible the ice clouds move equatorward and are carried out of the polar region as the springtime recession begins.

*Probable $H_2O$ surface deposits after $L_s$=200*. Water ice detections near 75°S, 250°E after $L_s$=200 are likely surface deposits, for four reasons – 1.) they do not show low $H_2O$ ice grain indexes as per Eq. 4, therefore they are not likely to contain fine-grained water ice, 2.) they persist in the same location from $L_s$=205-225, 3.) this area is associated with the western edge of the 105km diameter Lau crater (74.2°S,252.2°E) and 4.) the MARCI images (particularly the $L_s$=211 image) show a low albedo patch on the steep western side of Lau. In addition, water ice detections at 84°S, 151°E and 85°S, 60°E are also candidates to be surface deposits since they lie around the edge of the SPRC plateau. Due to large grained $CO_2$ ice around the SPRC plateau, the $H_2O$ grain size index (Eq. 4) cannot be employed to determine the $H_2O$ grain size. These $H_2O$ deposits have disappeared by $L_s$=230-240.

*Probable $H_2O$ clouds after $L_s$=200.* Spatially coherent water ice detections near 70-75°S from 90-210°E persist over $L_s$=200-225, and are probably vestiges of the cloud system that is dominant prior to $L_s$=200, since some display low $H_2O$ ice grain size indexes (Eq. 4). It should be noted that MARCI images show no visible evidence for ice clouds, suggesting the clouds are tenuous.

*Cryptic phase ($L_s$=200-240).* Figures 3.2-3.3 cover this phase. The MARCI images clearly show the rapid appearance of the Cryptic Region from $L_s$=195-210, and the CRISM ice maps show evidence of pure water ice appearing in the Cryptic Region at that time. The water ice that appears in the Cryptic Region can be seen to follow topography, this is especially clear in $L_s$=209-216 when the rim of Ultimum Chasma (at 81°S, 150°E) is outlined by pixels where only $H_2O$ ice is detected. There is no clear indication of low $H_2O$ ice grain size index (Eq. 4) associated with the $H_2O$ ice spectral signatures in the Cryptic Region from $L_s$=200-240, therefore they are likely to be due to surface water ice deposits as suggested by some researchers (Bibring et al., 2004; Titus, 2005b). Promethei Chasma (above Ultimum Chasma in the polar stereographic representation in Figure 3) shows a mixture of $CO_2$ and $H_2O$ ice signatures at the same time period.

Most of the water ice in the Cryptic Region disappears during the $L_s$=200-240 time period, and it is around this time that there is a peak of 25 pr-$\mu$m and an average around 20pr-$\mu$m (from a baseline of ~5pr-$\mu$m) in water vapor in the southern latitudes (Smith et al., this issue). It seems reasonable to suggest the water ice from the Cryptic Region is the source of this spike in atmospheric water vapor. Alternately, other researchers have suggested $H_2O$ ice beneath the SPRC plateau may be the source of an increase in water vapor during this time (Titus et al., 2008).

The Cryptic Region is roughly 1.7 million km$^2$ at its greatest extent (using calculations based on the low albedo region in the MARCI image at $L_s$=211). Based on CRISM observations at $L_s$=209.4-216.1, roughly two-thirds of the Cryptic Region shows signs of water ice (~1.1 million km$^2$). If all of the water in the south polar peak in the atmosphere was sourced from the Cryptic Region, and if the area covered by the atmospheric peak is roughly 17 million km$^2$ (this assumes the above-baseline $H_2O$ vapor region covers 35°S-75°S degrees in latitude), and assuming the water was sourced from a homogenous, pure, single layered deposit, then the layer of water ice would be roughly 230 microns ((20-5)*(17/1.1)) thick. This is obviously a simplification of the true situation, but it gives a testable hypothesis for future models and observations.

*Asymmetric Retraction phase ($L_s$=240-290)*. Figures 3.3-3.4 cover this phase. The bright albedo feature (the "Mountains of Mitchel") at 70°S, 30°E is clearly visible in the MARCI images from $L_s$=230-270, and although CRISM coverage extends definitively over this feature only at $L_s$=261-267, it clearly shows the signature of fine grained $CO_2$ ice.

This phase of the cap retreat is almost entirely devoid of water ice detections. The last vestiges of water ice are detected at $L_s$=243-251 at 83°S, 125°E. This location is at the end of Promethei and Ultimum Chasmata and is a hot spot of water activity after the retraction of water ice clouds at $L_s$=205-250. It was also recognized in MY 27 by OMEGA and disappeared at $L_s$=254 (paragraph [58] of Langevin et al., 2007). No water ice deposits are detected near 156°E, 86.4°S at $L_s$=270, as reported by Titus (2005b), however this may be due to lack of coverage or interannual variability. The absence of ice lags at $L_s$=270 may be due to 1.) subpixel mixing of remnant $CO_2$ ice and dust (Titus et al., 2008), or 2.) elevated atmospheric temperatures or surface obscuration caused by the planet encircling dust event which occurred around $L_s$=270 in MY28.

*Stable phase ($L_s$=290-340)*. Figures 3.5-3.6 cover this phase. The most extensive CRISM coverage of the SPRC was collected from $L_s$=312-320. It is difficult to make out at the resolution of Figure 3, but from $L_s$=312-342 CRISM detects small deposits of water ice from 270-360°E, in much the same location as identified from THEMIS data and shown in Figure 3 of Piqueux et al. (2008). Prior to $L_s$=312, only $CO_2$ ice is detected in these locations.

*Recondensation phase ($L_s$=340-040)*. Figures 3.6-3.7 cover this phase. CRISM and MARCI coverage for this time is spotty due to low solar illumination conditions at this time. The $CO_2$ index map indicates that $CO_2$ has reappeared and extends further equatorward than 80°S at $L_s$=4-10 approximately symmetrically around the geographic pole. GCM simulations predict the condensation of $CO_2$ ice to begin around $L_s$=0-10 (Haberle et al., 2004), as the cap cools at the start of southern fall - this is in accord with our observations here. MOLA derived $CO_2$ deposition estimates also suggest symmetric deposition around the geographic pole (Smith et al., 2009), which is also supported by our observations. MARCI images are more ambiguous and do not show obvious signs of the deposition of $CO_2$ ice, due probably due to the lack of coverage over the pole itself and the very thin nature of the deposit at this time. CRISM observations show no difference in the grain size or albedo of the Cryptic Region up to $L_s$=16.2, therefore it is possible that whatever process differentiates the Cryptic Region from the rest of the $CO_2$ seasonal cap has not manifested itself by $L_s$=16.

We investigated several CRISM MSP images at this time and found evidence of the $CO_2$ condensation process. Figure 8 shows part of MSP 00009624_01, centered at 70°E, 67°S and obtained at $L_s$=11.7. This image shows an increase in albedo over the newly condensing cap, and typical ice cap features such as wind blown dark areas where the $CO_2$ ice has been stripped from some areas. An example spectrum of the $CO_2$ ice cap is also shown - the spectra of the high albedo areas show fine-grained $CO_2$ ice, as determined by the shallow absorption features at 2.28μm.

*Comparison to TES observations of MY 23 recession.* Kieffer et al. (2000) showed that limited thermal infrared coverage by TES over the Cryptic Region at $L_s$=12 and over the geographic pole at $L_s$=25 was not able to detect the recondensation of fine grained $CO_2$ frost. Kieffer et al. postulated that transparent slab ice was forming at that time, based on the nearly constant brightness temperature over the 250-500cm$^{-1}$ spectral range of an average of 60 TES spectra. However, CRISM observations during MY28 show evidence for fine grained ice, and in addition no differences in grain size are detected over the Cryptic Region (see in particular the observations of 1.435 μm depth at $L_s$=4.6-10.9). This result is difficult to reconcile with TES observations, although one possibility is that the $CO_2$ ice coverage is sufficiently thin at this stage that its effect on the 250-500 cm$^{-1}$ region is negligible and TES is mostly sensing the soil beneath the ice. The nature of the early $CO_2$ condensation process clearly requires further study.

*Comparisons to THEMIS high resolution atmospheric observations.* Inada et al. (2007) used high resolution THEMIS images to observe 'trough clouds' with heights of around 600m in topographic lows around the south polar cap from $L_s$=251-320 in MY26. No detections of water ice clouds were made in trough regions in the Ls=251-320 by CRISM in MY28. Inada et al. do not have data covering periods earlier than $L_s$=240. Although it is possible some pre-$L_s$=251 water ice detections in topographic lows may be associated with fogs or ground-hugging clouds, most topographic water ice detections are not associated with low $H_2O$ ice grain size index ratios (Eq. 4) making surface deposits a more likely explanation for these water ice signatures.

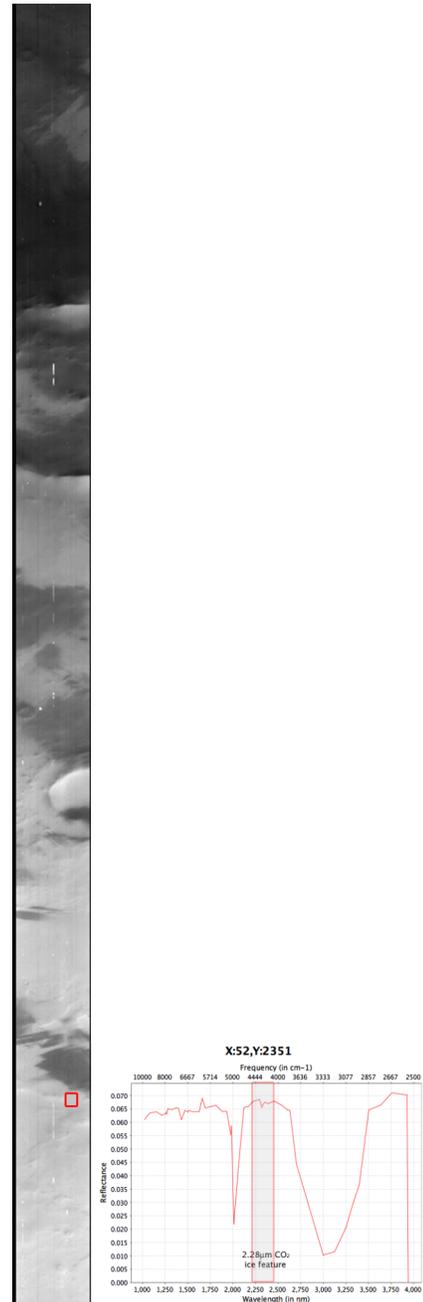

Figure 8. Evidence for $CO_2$ seasonal cap recondensation at $L_s$=11.7. (left) part of image MSP00009624_01, centered at 70E, 67S. The image shows the spectral band at 1.079μm. (right) MSP spectrum from pixel at x=52, y=2351 showing $CO_2$ absorption features at 2.28 μm. The image is 10km wide and north is up.

*Comparison to OMEGA observations of MY27 recession*. OMEGA detected water ice clouds over Hellas at $L_s$=140-154 (Figure 12c of Langevin et al., 2007) but due to the elliptical orbit of Mars Express, OMEGA did not return to observe the south polar cap until $L_s$=183.9-192.5, when it detected low amounts of water ice over the seasonal cap, with a 'dry spoke' along 60°E longitude, in the direction of Hellas. Our results expand upon this OMEGA observation, bringing the detection of spatially coherent water ice clouds to earlier in the year - at least to $L_s$=167.9 and perhaps earlier to $L_s$=161.9 (although coverage at $L_s$=161.9 is sparse). Our observations also place a large reduction in water ice cloud coverage at $L_s$=200-204, and further, we resolve the clouds as extending off the cap at latitudes of ~56-58°S between longitudes 90-150°E from $L_s$=167-199, into the Promethei Terra region south east of Hellas.

**Interpretation of $CO_2$ Grain Size Results**

Figures 7.1-7.5 show the histograms for derived seasonal $CO_2$ ice grain sizes. Although the patchy coverage of the polar regions frustrates definitive "pole-wide" knowledge about grain size estimates, we make the assumption that our observations are representative of the rest of the polar cap at the time of observation.

Despite the limitations of our $CO_2$ ice grain size modeling (which assumes 100% coverage of pure $CO_2$ ice with no dust cover, leading to 'minimum apparent' grain size estimates), the histograms show that the range of grain sizes across the $CO_2$ ice cap is relatively restricted according to the phase the cap is in (at least at the ~187.5m spatial scale).

*Pre-sublimation $CO_2$ ice grain sizes*. From $L_s$=162-208, the range of relative 2.28 $\mu$m band depths of $CO_2$ ice is at its widest. During this period, the largest relative band depth of 0.535 are achieved at Ls=191-199. Assuming they are due to $CO_2$ alone and our models are correct, this would correspond to minimum apparent grain sizes slightly less than 7±1cm.

*Cryptic period $CO_2$ ice grain sizes*. As spring begins to warm up the south polar region (from $L_s$=209-267) the grain sizes adopt a more restricted profile, becoming smaller (the largest band depths now range from 0.3-0.4 relative band depth, corresponding to a grain size of 2-3cm), and with a peak just short of 0.2 relative band depth (suggesting most of the $CO_2$ ice grains sizes cluster around an average of 5-10±1mm).

*Asymmetric retraction period $CO_2$ ice grain sizes*. During $L_s$=271-303, relative bands depth profiles for $CO_2$ ice continue to constrict. The maximum band depths of 0.25 (corresponding to a grain size of 10±1mm) and most of the $CO_2$ ice relative band depths are around 0.1-0.2 (corresponding to a grain sizes of 2.5-7.5±1mm).

*Stable period $CO_2$ ice grain sizes*. During the summer stable period, the relative band depths are also the most stable and constricted. Practically all of the relative band depths are below 0.2 (grain size of 7.5±1mm) and most cluster around the 0.1-0.15 relative band depth (corresponding to a grain size of 2.5-5±1mm).

*Recondensation period $CO_2$ ice grain sizes*. During the recondensation period, the relative band depths are also extremely stable, but a prominent histogram peak is not seen above the threshold (the highest number of pixels have grain sizes at the threshold, suggesting the peak is below the threshold and therefore not shown on the histogram). This is most likely because the majority of $CO_2$ ice grain sizes are below our threshold (0.07 relative band depth, corresponding to 1.2mm $CO_2$ ice grains, according to our spectral models). The lack of larger grained ice is directly related to the fact that the SPRC is not imaged in these sequences, and only fine-grained $CO_2$ from the condensing cap is detected. The small number of pixels that are identified as $CO_2$ ice makes the histograms more sensitive to noisy pixels, and they display a long tail of relative band depths. We interpret these histograms as showing no prominent mode of grain sizes and suggest that most of the $CO_2$ ice grain sizes are below 1.2mm at this time.

*Interpretations of decreasing of $CO_2$ ice grain sizes*. During the $L_s$=200-310 springtime period, the SPRC is covered by our observations and the maximum grain sizes of the residual polar cap are clearly decreasing. This observation might be explained in two ways: 1.) by relatively small subliming grains of $CO_2$ ice being transported poleward and getting cold trapped on the SPRC as fine-grained deposits (Kieffer et al., 2000) or 2.) by $CO_2$ ice grains sublimating and decreasing in size in place on the SPRC. Both options could explain our observation, and this process requires further study.

One interpretation of the peaks in CO2 ice grain size histograms (combined with $CO_2$ abundance maps in Figure 3) is that the SPRC $CO_2$ ice is responsible for the maximum relative band depths in each histogram. In early pre-sublimation CRISM observations the SPRC $CO_2$ ice grains experience growth conditions until $L_s$=190 when grain sizes of ~7cm are reached, and over summer these grains lose mass/sublime or are covered by finer grains until they display a stable size during summer ($L_s$=310) of ~5mm.

*Comparison with previous grain size results*. The first grain size estimations for $CO_2$ in the south pole of Mars came from studies of the Mariner 7 IRS dataset covering the SPRC at $L_s$=200. This study suggested quite large (millimeter to centimeter sized) grains (Calvin, 1990; Calvin and Martin, 1994).

OMEGA observations at $L_s$=142 uncovered "slab ice" at 59S, 344E, which was modeled to indicate grain sizes of 30cm (Langevin et al., 2007). CRISM did not observe this region at $L_s$=142. We found that $CO_2$ grain sizes were larger on the SPRC than on the seasonal cap, which was also reported by Kieffer et al. (2000) and Langevin et al. (2007). Our minimum apparent grains size estimates of ~7±1cm for grains on the SPRC at $L_s$=190 are similar to OMEGA (5-10cm, paragraph [39] and [58] of Langevin et al., 2007 and page 17 of Doute et al., 2007 - at $L_s$=335-348) and TES (~10cm, $L_s$=193 on perennial cap, Figure 3 caption of Kieffer et al., 2000). Kieffer also reports that the grain size decreases by the summer ($L_s$=316, Figure 3b of Kieffer et al., 2000). This is in broad accordance with our observations.

The results reported here for CRISM data are in reasonable agreement with those of Glenar et al., (2005) who found telescopic spectra of the bright regions of the south polar cap at $L_s$=231 were well modeled by $CO_2$ ice with a grain size of 13mm (this is just above our average grain size of 5-10±1mm for this period), and Hansen et al., (2005) who suggested PFS observations of the SPRC at $L_s$=330 were well modeled by a surface containing $CO_2$ ice of grain size 5-10mm (this is just above our average of 2.5-5±1mm for the stable phase).

The grain size estimates presented here are in broad accordance with those of Doute et al. (2006) and Langevin et al. (2007) despite the fact that we have used a different grain size estimate method, and we have made no assumptions about contamination by dust or water ice. All of the Doute et al. models of the SPRC use 400-1600ppm $H_2O$ with 300-400 $\mu$m grain size, which could be appropriate for late summer observations.

*$CO_2$ slab ice in the Cryptic Region*. CRISM data provide no evidence for $CO_2$ slab ice, or any large grained $CO_2$ ice (no relative band depths for 2.28$\mu$m band of greater than ~0.16 corresponding to grain sizes of ~5mm) in the Cryptic Region, in agreement with the results reported by the OMEGA team (Langevin et al., 2007). In fact, the Cryptic Region, even before it shows Cryptic behavior (before $L_s$=190) is characterized by an absence of $CO_2$ features. However, $CO_2$ ice clearly controls the temperature of the surface, as determined by the thermal infrared observations of TES (Kieffer et al., 2000). One explanation is that a layer of dust on the surface of the Cryptic Region lies on top of the $CO_2$ ice slab, in thermal equilibrium with $CO_2$ ice beneath (Titus et al., 2008). Langevin et al. presented spectral models that require the dust to be in the top 2mm of the surface (Langevin et al., 2006). The dust must be heavy enough to mask the $CO_2$ slab from detection in the VNIR, however it must be thin or discontinuous enough to allow radiation to penetrate and power the Cryptic Region cold jets, as has been proposed in Kieffer's standard cold jet model (Kieffer et al., 2006). An additional explanation for the lack of large path length $CO_2$ absorption features is that all photons in VNIR wavelengths entering the transparent $CO_2$ ice slab would be absorbed, leaving the only signature returned to the sensor that due to the top layer of dust.

# CONCLUSIONS

We have presented the results of CRISM and MARCI mapping during the first Mars year of Mars Reconnaissance Orbiter science mapping operations.

The observations reported herein have for the first time:
1.) mapped latitudinal/longitudinal extent of spatially coherent water ice clouds above the polar cap from $L_s$=160-200. These clouds may be used as tracers of atmospheric dynamics, including possible indications of east-west asymmetry in weather patterns (Brown et al., 2008b).
2.) within the Cryptic Region, uncovered the presence of $H_2O$ ice with no accompanying $CO_2$ ice signatures. These identifications are an intriguing part of the Cryptic Region puzzle. If ice sourced from the Cryptic Region is solely responsible for the summertime rise in water vapor in the southern hemisphere, Cryptic Region water ice deposits may form a layer more than 200 microns thick.
3.) within the Cryptic Region, mapped areas at the ~187.5m/pixel scale where no $CO_2$ or $H_2O$ ice was observed (due either to obscuration by dust or sublimation of the icy volatiles).
4.) mapped the seasonal variations in the strength of the 2.28 $\mu$m $CO_2$ absorption relative band depth, and inferred that $CO_2$ ice grains and apparent grain sizes of up to ~7±1cm are present prior to the springtime sublimation of the south polar cap, compared with average grain sizes of 2.5-5±1mm in the stable summer time residual polar cap.
5.) mapped the recondensation of the $CO_2$ seasonal cap from $L_s$=0-30 and provided evidence of a very thin, fine-grained deposit that extends beyond 80°S by $L_s$=4-10. The deposit is roughly symmetrical around the geographic pole. Whatever process is responsible for forming the Cryptic Region is not apparent before $L_s$=16, since no difference in grain size or albedo (from the rest of the condensing cap) is observed in that region up to that time.

# ACKNOWLEDGEMENTS

We would like to thank the entire CRISM Team, particularly the Science Operations team at JHU APL, and also the MARCI PI (Mike Malin) and his staff. We thank Sylvain Piqueux and an anonymous reviewer for their helpful guidance. We thank Ted Roush for supplying optical constants for palagonite. Special thanks to Tim Titus, Mike Wolff, Yves Langevin and Todd Clancy for illuminating discussions. This investigation was partially funded by NASA Grant NNX08AL09G.